\documentclass[12pt,aps,prd,superscriptaddress,showpacs,longbibliography,floatfix,nofootinbib]{revtex4-1}

\usepackage[utf8]{inputenc}
\pdfoutput=1

\usepackage{color}
\usepackage{graphicx}   
\usepackage{bm}
\usepackage{amsmath}
\usepackage{amsfonts}
\usepackage{eufrak}
\usepackage{hyperref}

\def\sumint{\,\hbox{$\sum$}\!\!\!\!\!\!\!\int}

\begin{document}

\title{Simple non-perturbative resummation schemes beyond mean-field II: thermodynamics of scalar $\phi^4$ theory in 1+1 dimensions at arbitrary coupling}

\author{Paul Romatschke}
\affiliation{Department of Physics, University of Colorado, Boulder, Colorado 80309, USA}
\affiliation{Center for Theory of Quantum Matter, University of Colorado, Boulder, Colorado 80309, USA}

\begin{abstract}
  Recently, non-perturbative approximate solutions were presented that go beyond the well-known mean-field resummation. In this work, these non-perturbative approximations are used to calculate finite temperature equilibrium properties for scalar $\phi^4$ theory in two dimensions such as the pressure, entropy density and speed of sound. Unlike traditional approaches, it is found that results are well-behaved for arbitrary temperature/coupling strength, are independent of the choice of the renormalization scale $\bar\mu^2$, and are apparently converging as the resummation level is increased. Results also suggest the presence of a possible analytic cross-over from the high-temperature to the low-temperature regime based on the change in the thermal entropy density.
\end{abstract}

\maketitle

\section{Introduction}

Recently, I presented a sequence of non-perturbative approximate solutions for scalar $\phi^4$ theory for arbitrary interaction strength \cite{Romatschke:2019rjk}. These approximate solutions contain, but allow to systematically improve on, the familiar mean-field approximation. In this work, I consider $\phi^4$ field theory in two dimensions at finite temperature as a natural extension of the zero-temperature study done in Ref.~\cite{Romatschke:2019rjk}.

Finite temperature quantum field theory is a mature and well-established discipline \cite{Laine:2016hma}. At high temperature, naive perturbation theory breaks down because of infrared singularities. These difficulties are by now understood to be cured by resumming an infinite number of Feynman diagrams, generating an effective in-medium (thermal) mass. This ``Hard-Thermal-Loop'' (HTL) resummation \cite{Braaten:1989mz} has led to a very successful program for calculating properties of field theories at finite temperature and/or density, cf. Refs.~\cite{Taylor:1990ia,Braaten:1990az,Braaten:1991gm,Braaten:1991dd,Kelly:1994dh,Flechsig:1995ju,Moore:1997sn,Carrington:1997sq,Andersen:1999fw,Bodeker:1999gx,Blaizot:1999ap,Bolz:2000fu,Karsch:2000gi,Peshier:2000hx,Blaizot:2000fc,Blaizot:2001vr,Blaizot:2001nr,Rebhan:2003wn,Romatschke:2003ms,Kraemmer:2003gd,Andersen:2004fp,Rebhan:2004ur,Laine:2006ns,CaronHuot:2007gq,Rychkov:2007uq,Ghiglieri:2013gia,Haque:2014rua,Gorda:2018gpy}.

So why invest time into developing novel resummation schemes, given the apparent success of the HTL resummation program?

First, despite resumming an infinite number of Feynman diagrams, the HTL resummation scheme is not fully non-perturbative in the sense that HTL results do not exhibit a sensible strong-coupling limit. Second, the HTL resummation scheme does not easily incorporate the physics of transport which typically requires resummation of higher-order Feynman diagrams. Third, observables exhibit an unphysical dependence on the renormalization scale choice, which is a property inherited from perturbative truncations of the full theory.

This provides the motivation to consider the resummation schemes R0-R3 described in Ref.~\cite{Romatschke:2019rjk} to test if any of these issues arsing for the HTL resummation scheme can be improved on. For simplicity of presentation, I chose to ignore transport properties for the present work and only study equilibrium thermodynamics.

\section{Finite temperature pressure of scalar $\phi^4$ theory in 2d}

Let me consider the path integral formulation of $\phi^4$ theory in two Euclidean dimensions given by
\begin{equation}
  Z=\int {\cal D}\phi e^{-S}\,,\quad S= \int d^{2}X \left(\frac{1}{2}\partial_\mu \phi \partial_\mu \phi+\frac{m^2}{2}\phi^2+\lambda \phi^4\right)\,,\quad X=\left(\tau,x\right)\,,
\end{equation}
where $\lambda$ has mass dimension two and $m^2>0$ is assumed. The Euclidean time direction $\tau$ is compactified on a circle with radius $\beta\equiv T^{-1}$, where $T$ is the equilibrium temperature of the system. Introducing an auxiliary field $\zeta$, the path integral may be re-written as 
\begin{equation}
  \label{eq:zgen}
  Z=\sqrt{\frac{\beta V}{16 \lambda \pi}}\int d\zeta_0 e^{-\frac{\zeta_0^2 \beta V}{16 \lambda}}\int {\cal D}\phi {\cal D}\zeta^\prime e^{-S_0-S_I}\,,
\end{equation}
where $V$ is the ``volume'' of the Euclidean direction $x$ and
\begin{equation}
  S_0=\frac{1}{2}\int d^2X\left[ \partial_\mu \phi \partial_\mu \phi+m^2 \phi^2+ i \zeta_0 \phi^2+\frac{\zeta^{\prime 2}}{2\lambda}\right]\,,\quad
  S_I=i \int d^2X \zeta^\prime \phi^2\,,
\end{equation}
and $\zeta_0$ is the global zero mode of $\zeta$. The resummation schemes R0-R3 introduced in Ref.~\cite{Romatschke:2019rjk} correspond to different approximation levels (R0 the ``coarsest'' and R3 the ``finest'') of the partition function.

\subsection{R0-level}

In the R0 scheme, the term $S_I$ is dropped completely, and the partition function may be evaluated analytically as \cite{Romatschke:2019rjk}
\begin{equation}
  Z_{R0}=\sqrt{\frac{\beta V}{16 \lambda \pi}}\int d\zeta_0 e^{-\beta V \left[\frac{\zeta_0^2}{16 \lambda}+J\left(\sqrt{m^2+i \zeta_0}\right)\right]}\,,
\end{equation}
where $J(\alpha)=J_0(\alpha)+J_T(\alpha)$ and in 
$d=1-2\epsilon$ Euclidean space dimensions \cite{Laine:2016hma}
\begin{equation}
  J_0(\alpha)=\frac{\alpha^2}{8 \pi \epsilon}+\frac{\alpha^2}{8\pi} \ln \frac{\bar \mu^2 e^1}{\alpha^2} + {\cal O}(\epsilon)\,,\quad J_T(\alpha)=-\frac{\alpha T}{ \pi} \sum_{n=1}^\infty \frac{K_1\left(\frac{ n \alpha}{T}\right)}{n}+{\cal O}(\epsilon)\,.
\end{equation}
Here $\bar \mu^2=4 \pi \mu^2 e^{-\gamma_E}$ is the $\overline{\rm MS}$  scale parameter that in finite-temperature field theory literature is customarily varied by a factor two around the first non-vanishing Matsubara frequency, e.g.
$\bar\mu \in \left[\pi T,4 \pi T\right]$. Physical observables are not meant to depend on $\bar\mu$, hence varying $\bar\mu$ in truncations of the full theory is used to test for the sensitivity of results to higher-order terms not considered in the approximation.

In the large volume limit $V\rightarrow \infty$, the partition function in the R0 approximation may be evaluated through a saddle-point approximation, finding
\begin{equation}
  Z_{R0}=e^{-\beta V \left[-\frac{z^{*2}_{R0}}{16 \lambda}+J\left(\sqrt{m^2+z_{R0}^*}\right)\right]}\,,\quad i \zeta_0=z^*_{R0}=4 \lambda I\left(\sqrt{m^2+z_{R0}^*}\right)\,,
\end{equation}
where $I(\alpha)=2 \frac{d J(\alpha)}{d\alpha^2}$. At zero temperature, the theory is renormalized by requiring a finite pole mass of the two-point function $\langle \phi(x)\phi(0\rangle$, which in the R0 approximation leads to $m^2_R=m^2+\frac{\lambda}{\pi \epsilon}$ \cite{Romatschke:2019rjk}. Solving the resulting renormalization group equation hence gives the running of the renormalized mass $m_R^2$ with the renormalization scale $\bar\mu^2$ as
\begin{equation}
  \label{eq:r0running}
  m_R^2\left(\bar\mu^2\right)=m_F^2-\frac{\lambda}{\pi}\ln \frac{\bar\mu^2}{m_F^2}\,,
  \end{equation}
where $m_F^2$ is the value of the renormalized mass at some fiducial scale. Given that $\lambda$ has mass dimension two, it is useful to consider units in which all other dimensionful quantities are expressed in terms of $\lambda$, e.g.
\begin{equation}
  \hat m_R^2\equiv \frac{m_R^2}{\lambda}\,,\quad \hat T\equiv \frac{T}{\sqrt{\lambda}},\quad \hat \mu^2=\frac{\bar \mu^2}{\lambda}\,.
  \end{equation}
Note that in these units, the weak-coupling regime corresponds to high temperature $\hat T\rightarrow \infty$, whereas low temperature corresponds to strong coupling, similar to studies of dimensionally reduced gauge theories and gauge/gravity duality \cite{Aharony:2004ig,Kawahara:2007ib,Hanada:2016qbz}. For simplicity of notation, I will drop the hat notation in the following, effectively using units where $\lambda=1$.

\begin{figure}[t]
  \includegraphics[width=\linewidth]{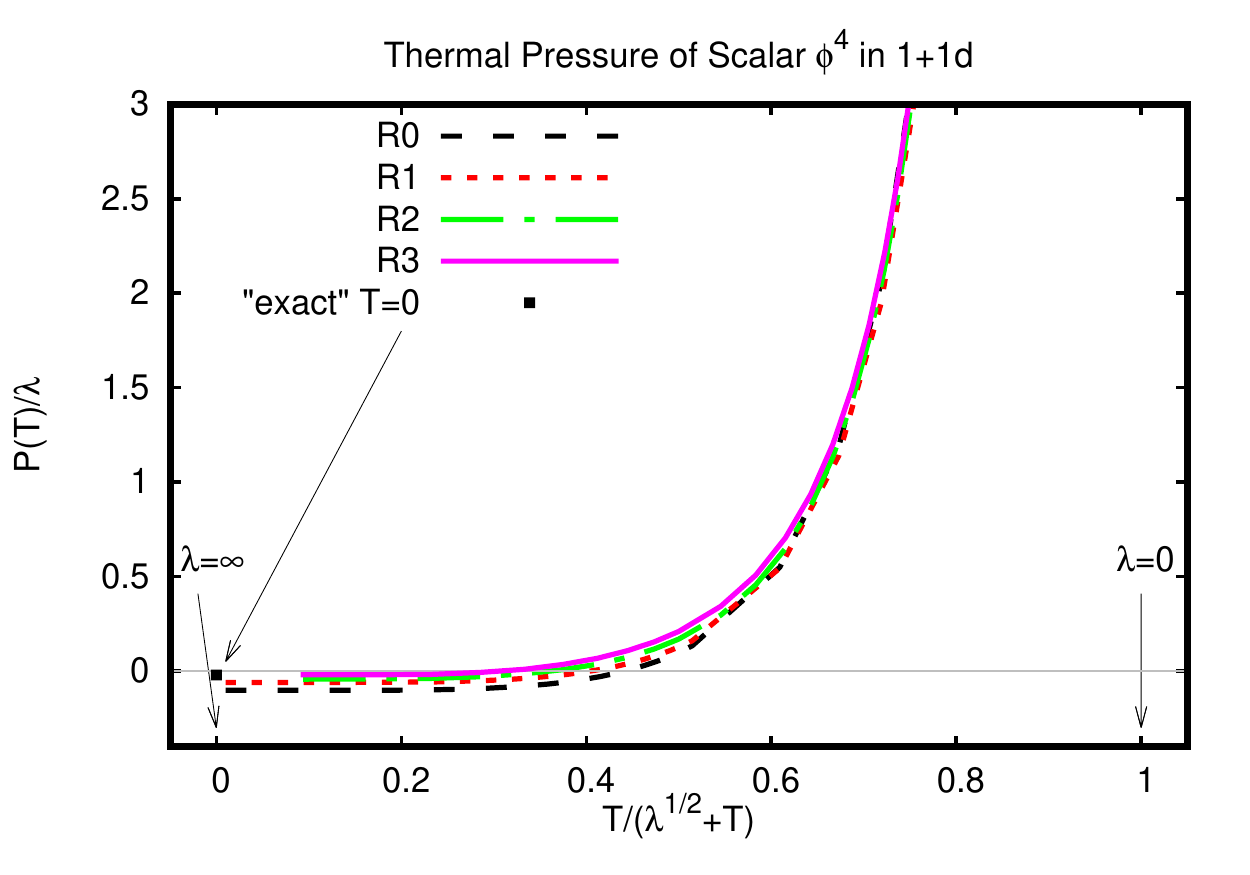}
  \caption{\label{fig1} Thermal pressure as a function of re-scaled temperature $T/\sqrt{\lambda}$ for $m_F^2=\lambda$ in the R0-R3 approximation scheme. Results do not depend on the renormalization scale choice $\bar\mu$. Horizontal plot axis is expressed as $\frac{T}{T+\sqrt{\lambda}}$ to allow compactification of the whole interval $T\in[0,\infty)$. Arrows indicate location of weak and strong coupling regime $\lambda=0$, $\lambda=\infty$, respectively, as well as the location of the zero-temperature result from high precision (``exact'') calculations (\ref{eq:t0exact}) using Refs.~\cite{Serone:2018gjo,Elias-Miro:2017xxf}.  Full line is a guide to the eye.} 
\end{figure}

The partition function can be written as
\begin{equation}
  \label{eq:zform}
  Z=e^{\beta V \left(P(T,\bar \mu^2)+P_{\rm div}\right)}\,,
\end{equation}
where $P_{\rm div}=\frac{m^4}{16}$ is a divergent contribution to the cosmological constant in the R0-level approximation and the finite-temperature pressure is given by
\begin{equation}
  \label{eq:pr0}
  P_{R0}(T,\bar \mu^2)=- J_T(M)+\frac{M^4}{16}-\frac{M^2 m_R^2}{8}-\frac{M^2}{8 \pi}\ln \frac{\bar \mu^2 e^1}{ M^2}\,,
\end{equation}
with a self-consistent pole mass $M$ determined by the solution of the ``gap'' equation
\begin{equation}
  \label{eq:r0gap}
  M^2=m^2+z_{R0}^*=m_R^2+\frac{1}{\pi}\ln \frac{\bar \mu^2}{ M^2}+4 I_T(M)=m_F^2+\frac{1}{\pi}\ln\frac{m_F^2}{M^2}+4 I_T(M)\,.
  \end{equation}
Using the running of $m_R^2$  in the R0-level approximation (\ref{eq:r0running}), one may verify explicitly that the finite-temperature pressure $P_{R0}(T,\bar \mu^2)$ is independent from the choice of the renormalization scale $\bar\mu^2$. At very high temperature, the R0-pressure reduces to the pressure of a free scalar field in two dimensions,
\begin{equation}
  P_{free}=\frac{\pi T^2}{6}\,,
\end{equation}
as expected for a weakly coupled field theory. At finite temperature, the R0 level approximation results in a reduction of the pressure from the free result, which contains all orders in perturbation theory \textit{partially}\footnote{It is worth recalling that the R0 level approximation corresponds to the leading $1/N$ result from the N-component scalar field theory in the limit $N\rightarrow \infty$.} through the self-consistent solution of the gap equation (\ref{eq:r0gap}). At zero temperature, the $P_{R0}(T=0)=-\frac{m_F^4}{16}-\frac{m_F^2}{8\pi}$ contains a finite contribution to the cosmological constant\footnote{To avoid confusion, re-instating powers of $\lambda$, the zero temperature contribution $-\frac{m_F^4}{16\lambda}-\frac{m_F^2}{8\pi}$ has been subtracted in Refs.~\cite{Romatschke:2019rjk,Serone:2018gjo,Elias-Miro:2017xxf} when requiring the cosmological constant to vanish at $\lambda=0$. Thus, the renormalization condition adopted in this work differs from these references.}. A plot of the pressure for the R0 approximation is shown in Fig.~\ref{fig1} for $m_F^2=1$. While only part of the temperature range is visible in this figure, the pressure is well-behaved for all temperatures, and smoothly interpolates from the weak-coupling, high-temperature regime to the strong-coupling, low temperature regime.

\subsection{R1-level}

Without further input, it is not clear how good an approximation the R0-level resummation for the true finite-temperature pressure of scalar $\phi^4$ really is. A step forward can be made by considering the next best approximation level, R1, which arises from (\ref{eq:zgen}) by a suitable re-shuffling of terms between $S_0,S_I$ (see Ref.~\cite{Romatschke:2019rjk} for details), finding
\begin{equation}
  Z_{R1}=\sqrt{\frac{\beta V}{16 \pi}}\int d\zeta_0 e^{-\beta V \left[\frac{\zeta_0^2}{16 }+J\left(\sqrt{m^2+3 i \zeta_0}\right)-2 I^2\left(\sqrt{m^2+3 i \zeta_0}\right)\right]}\,.
\end{equation}
In the large volume limit, the partition function can once again be evaluated through a saddle point approximation, finding $i \zeta_0=z_{R1}^*=4 I\left(\sqrt{m^2+3 z_{R1}^*}\right)$. The zero-temperature pole mass is rendered finite by introducing a renormalized mass squared $m_R^2=m^2+\frac{3}{\pi \epsilon}$ \cite{Romatschke:2019rjk}, which leads to the mass running as
\begin{equation}
  \label{eq:r1running}
  m_R^2\left(\bar\mu^2\right)=m_F^2-\frac{3}{\pi}\ln\frac{\bar\mu^2}{m_F^2}\,,
  \end{equation}
cf. the corresponding equation (\ref{eq:r0running}) in the R0-level approximation. The partition function can once again be written in the form (\ref{eq:zform}) with a divergent contribution $P_{\rm div}=\frac{m^4}{48}$ and a finite-temperature pressure
\begin{equation}
  \label{eq:pr1}
  P_{R1}(T,\bar\mu^2)=-J_T(M)+\frac{M^4}{48}-\frac{M^2 m_R^2}{24}-\frac{M^2}{8\pi}\ln \frac{\bar\mu^2 e^1}{M^2}\,.
\end{equation}
Here $M^2$ is the self-consistent pole mass determined as the solution of the gap equation
\begin{equation}
  \label{eq:r1gap}
  M^2=m^2+3 z_{R1}^*=m_R^2+\frac{3}{\pi}\ln \frac{\bar \mu^2}{M^2}+12 I_T(M)=m_F^2+\frac{3}{\pi}\ln\frac{m_F^2}{M^2}+12 I_T(M)\,.
\end{equation}
Similar to what was found for the R0-level approximation, the running $m_R^2$ ensures that the R1-level pressure (\ref{eq:pr1}) is independent from the choice of the renormalization scale $\bar\mu^2$. (This is somewhat trivial for a super-renormalizable theory such as $\phi^4$ in 1+1 dimensions. However, the 
behavior persists for theories that are just renormalizable, as has been explicitly shown in Ref.~\cite{Blaizot:2000fc} corresponding to the R1 scheme for $\phi^4$ theory in 3+1 dimensions). Results for $P_{R1}(T)$ are shown in Fig.~\ref{fig1} as a function of temperature. While the leading perturbative correction term to $P_{free}$ originating from $P_{R0}(T)$ is only a third of that from $P_{R1}(T)$, Fig.~\ref{fig1} shows that R0 and R1-level approximations give similar results for the overall pressure magnitude for all values of temperatures/couplings shown. (There are notable relative differences for low temperatures, given that $P_{R1}(T=0)=-\frac{m_F^4}{48}-\frac{m_F^2}{8\pi}$ whereas $P_{R0}(T=0)=-\frac{m_F^4}{16}-\frac{m_F^2}{8\pi}$.)

\subsection{R2-level}

While both the R0 and R1-level approximations are non-perturbative in character,they correspond to mean-field-type resummations in the sense that only in-medium mass terms, but no in-medium thermal widths, are generated. Therefore, since the physics of thermal widths is not included in the R0, R1 approximations, one might worry that results based on R0, R1, despite being close to each other, could be far from the true, physical result. This indeed happens for the zero-temperature case where finite mass terms generated by R0, R1 can be renormalized away \cite{Romatschke:2019rjk}, and qualitatively different results are found for the R2, R3 level approximations.

For these reasons, it is important to study the R2-level approximation that includes the physics of thermal widths non-perturbatively. Rewriting of the terms 
$S_0,S_I$ in (\ref{eq:zgen}) by introducing dynamic propagators for both the $\phi$ and $\zeta^\prime$ fields (see Ref.~\cite{Romatschke:2019rjk} for details), one finds
\begin{eqnarray}
  Z_{R2}&=&\sqrt{\frac{\beta V}{16 \pi}}\int d\zeta_0 e^{-\beta V S^{\rm eff}_{R2}[i\zeta_0]}\,,\nonumber\\
  S^{\rm eff}_{R2}[i\zeta_0]&=&\frac{\zeta_0^2}{16}+\frac{1}{2}\sumint_K\ln \left(K^2+m^2+i\zeta_0+\Pi(K)\right)-\frac{1}{2}\sumint_K \Pi(K) G(K)\nonumber\\
  &&+\frac{1}{2}\sumint_K\ln \left(1+2 \Sigma(K)\right)-\frac{1}{2}\sumint_K \Sigma(K) D(K)+\vcenter{\!\hbox{\includegraphics[width=0.12\linewidth]{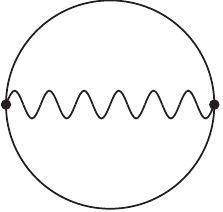}}}\,,
\end{eqnarray}
where the $\phi$-field propagator $G(K)=\left[K^2+m^2+i\zeta_0+\Pi(K)\right]^{-1}$ is denoted by a straight line, and the $\zeta^\prime$ field propagator $D(K)=2 \left[1+2 \Sigma(K)\right]^{-1}$ is denoted by a wiggly line. Furthermore I use $K=\left(\omega_n,k\right)$ to denote the Euclidean two-momentum where $\omega_n=2\pi n T$ are the bosonic Matsubara frequencies, and
\begin{equation}
  \sumint_K\equiv  \bar \mu^{2\epsilon}(4\pi)^{2 \epsilon}e^{\gamma \epsilon} T \sum_{\omega_n}\int \frac{d^{1-2 \epsilon}k}{(2 \pi)^{1-2\epsilon}} \,,
  \end{equation}
to denote the thermal sums and integrals in 1+1 dimensions (cf. Ref.~\cite{Laine:2016hma}). The self-energies $\Pi(K),\Sigma(K)$ are fixed by requiring that first non-trivial corrections arising from $\zeta^\prime$ in $S_I$ cancel when calculating the two-point functions $\langle \phi(X)\phi(0)\rangle$, $\langle\zeta^\prime(X)\zeta^\prime(0)\rangle$. This results in \cite{Romatschke:2019rjk}
\begin{equation}
  \label{eq:r2pis}
  \Pi(P)=4\ \vcenter{\!\hbox{\includegraphics[width=0.12\linewidth]{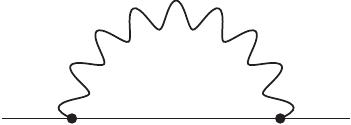}}}=4 \sumint_K D(K) G(P-K)\,,\quad
  \Sigma=2\ \vcenter{\!\hbox{\includegraphics[width=0.12\linewidth]{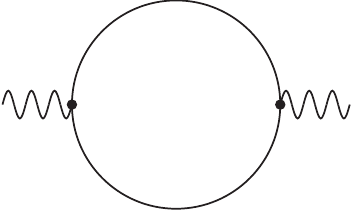}}}=2\sumint_K G(K) G(P-K)\,.
\end{equation}
In the large volume limit, the partition function $Z_{R2}$ can once again be evaluated through a saddle point approximation, finding
\begin{equation}
  i \zeta_0=z_{R2}^*=4 \sumint_K G(K)\,.
  \end{equation}
The zero-temperature inverse propagator $G^{-1}(K)$ is rendered finite by the same renormalization prescription that was used for R1, e.g. $m_R^2=m^2+\frac{3}{\pi \epsilon}$ \cite{Romatschke:2019rjk}, which leads to the mass running as in Eq.~(\ref{eq:r1running}). One thus finds
\begin{equation}
  G^{-1}(K)=K^2+m^2+3 z_{R2}^*-\delta \Pi(K)=K^2+m_F^2+\delta m^2-\delta \Pi(K)\,,
\end{equation}
where $\Pi(K)=2 z^*-\delta \Pi(K)$ and
\begin{eqnarray}
  \delta m^2&=&\frac{3}{\pi}\ln \frac{m_F^2}{m_F^2+\delta m^2}+12 I_T\left(\sqrt{m_F^2+\delta m^2}\right)+12 \sumint_K G(K)\frac{\delta \Pi(K)}{K^2+m_F^2+\delta m^2}\,,\nonumber\\
  \delta \Pi(K)&=&8 \sumint_Q G(K-Q)\frac{2 \Sigma(Q)}{1+2 \Sigma(Q)}\,.
\end{eqnarray}
Noting the cancellation
\begin{equation}
  \label{eq:r2rel}
  -\frac{1}{2}\sumint_K \Sigma(K) D(K)+\vcenter{\!\hbox{\includegraphics[width=0.12\linewidth]{one-loopR2}}}=0\,,
\end{equation}
the partition function can be written in the form (\ref{eq:zform}) with a divergent contribution $P_{\rm div}=\frac{m^4}{48}$   
 and a finite-temperature pressure given by
\begin{eqnarray}
  \label{eq:pr2}
  P_{R2}(T,\bar\mu^2)&=&-\frac{1}{2}\sumint_K \ln \frac{K^2+m_F^2+\delta m^2-\delta \Pi(K)}{K^2+m_F^2+\delta m^2}-\frac{1}{4}\sumint_K \delta \Pi(K) G(K)\nonumber\\
  &&-\frac{1}{2}\sumint_K \ln \left[1+2 \Sigma(K)\right]+\frac{1}{2}\sumint_K \Sigma(K) D(K)\nonumber\\
  &&-J_T\left(\sqrt{m_F^2+\delta m^2}\right)+\frac{(m_F^2+\delta m^2)^2}{48}\nonumber\\
  &&-\frac{(m_F^2+\delta m^2)}{24}\left(m_F^2+\frac{3}{\pi}\ln \frac{m_F^2 e^1}{m_F^2+\delta m^2}\right)\,.
%
\end{eqnarray}
Note that again, the dependence on the renormalization scale $\bar \mu^2$ has dropped out in $P_{R2}$.

Results for $P_{R2}(T)$ can be obtained numerically using the same methods as those described in the appendix of Ref.~\cite{Romatschke:2019rjk}. The only difference with respect to the algorithm described in Ref.~\cite{Romatschke:2019rjk} is that I explicitly evaluate the sum over Matsubara frequencies instead of performing a continuum integral. This approach becomes numerically expensive for small temperatures, but I find that for $T\gtrsim 0.1$ acceptable numerical accuracy can be obtained using only the first one hundred Matsubara frequencies. (The numerical code is publicly available at \cite{codedown}).

Numerical results obtained in this manner for $P_{R2}(T)$ are compared to $P_{R0}(T), P_{R1}(T)$ in Fig.~\ref{fig1}. As can be seen from this figure, the R2-level results for the pressure are in good quantitative agreement with the R0 and R1 results for all values of the temperature/coupling shown. Even the zero-temperature limit $P_{R2}(T=0)\simeq -0.0433$ is quantitatively similar to the results found for the R1-level approximation\footnote{As a non-trivial check on the numerics, note that converting to the renormalization scheme chosen in Ref.~\cite{Romatschke:2019rjk} one finds $-P_{R2}(T=0)-\frac{m_F^4}{48}-\frac{m_F^2}{8\pi}\simeq-0.017$, matching the result for the vacuum energy in the R2 approximation at $g=\frac{m_F^2}{\lambda}=1$ in Ref.~\cite{Romatschke:2019rjk}.}. This strongly suggests that the overall magnitude of the thermal pressure is dominated by physics arising from the non-perturbative mass-resummation, with contributions from thermal widths being quantitatively sub-leading.


\subsection{R3-level}

Increasing the resummation level further, one obtains the R3-level scheme where
\cite{Romatschke:2019rjk}
\begin{eqnarray}
  Z_{R3}&=&\sqrt{\frac{\beta V}{16 \pi}}\int d\zeta_0 e^{-\beta V S^{\rm eff}_{R3}[i\zeta_0]}\,,\nonumber\\
  S^{\rm eff}_{R3}[i\zeta_0]&=&\frac{\zeta_0^2}{16}+\frac{1}{2}\sumint_K\ln \left(K^2+m^2+i\zeta_0+\Pi(K)\right)-\frac{1}{2}\sumint_K \Pi(K) G(K)\nonumber\\
  &&+\frac{1}{2}\sumint_K\ln \left(1+2 \Sigma(K)\right) +2\ \vcenter{\!\hbox{\includegraphics[width=0.12\linewidth]{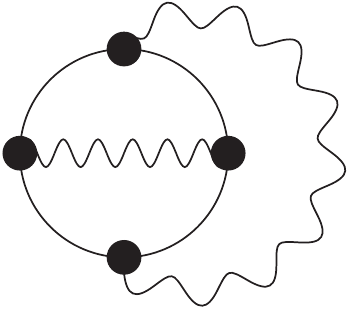}}}\,,
\end{eqnarray}
and where the R3-equivalent of (\ref{eq:r2rel}) was used. For the R3-level scheme, the self-energies are given by
\begin{equation}
  \label{eq:r3pis}
  \Pi(X)=4\ \vcenter{\!\hbox{\includegraphics[width=0.12\linewidth]{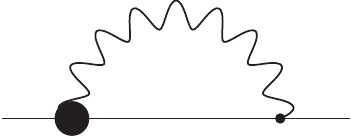}}}\,,\quad
  \Sigma(X)= 2\ \vcenter{\!\hbox{\includegraphics[width=0.12\linewidth]{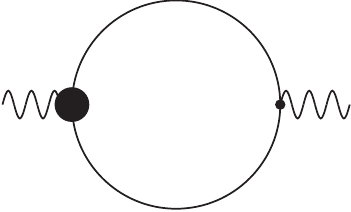}}}\,,
\end{equation}
where the resummed vertex $\Gamma=1+\delta \Gamma$ obeys
\begin{eqnarray}
  \label{eq:r3gamma}
  \vcenter{\!\hbox{\includegraphics[width=0.12\linewidth]{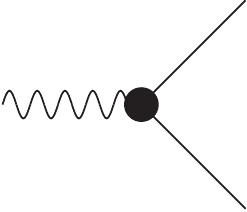}}}&=&1-4\vcenter{\!\hbox{\includegraphics[width=0.12\linewidth]{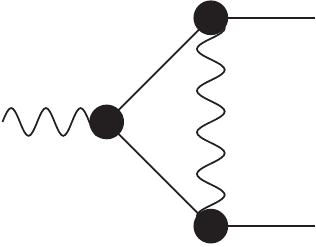}}}\,,\\
\end{eqnarray}
and is graphically represented as a ``blob''. As outlined in Ref.~\cite{Romatschke:2019rjk}, it is possible to recast $S_{R3}^{\rm eff}$ as sum over effective one-loop integrals by writing 
\begin{equation}
  \label{eq:deltaS}
  2\ \vcenter{\!\hbox{\includegraphics[width=0.12\linewidth]{one-loopR3}}} = -\frac{1}{4}\sumint_K D(K)  \delta\Sigma(K)\,,\quad
   \delta \Sigma(K) = 2\ \vcenter{\!\hbox{\includegraphics[width=0.12\linewidth]{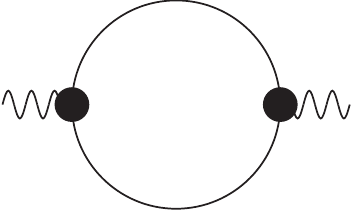}}}-2 \ \vcenter{\!\hbox{\includegraphics[width=0.12\linewidth]{one-loopSigmar3}}}\,.
\end{equation}
In the large volume limit, the partition function $Z_{R3}$ can once again be evaluated through a saddle point approximation, finding
\begin{equation}
  i \zeta_0=z_{R3}^*=4 \sumint_K G(K)\,.
  \end{equation}
The zero-temperature inverse propagator $G^{-1}(K)$ is rendered finite by the same renormalization prescription that was used for R1 and R2, e.g. $m_R^2=m^2+\frac{3}{\pi \epsilon}$, which leads to the mass running as in Eq.~(\ref{eq:r1running}). One thus finds
\begin{equation}
  G^{-1}(K)=K^2+m^2+3 z_{R3}^*-\delta \Pi(K)=K^2+m_F^2+\delta m^2-\delta \Pi(K)\,,
\end{equation}
where $\Pi(K)=2 z^*-\delta \Pi(K)$ and
\begin{eqnarray}
  \delta m^2&=&\frac{3}{\pi}\ln \frac{m_F^2}{m_F^2+\delta m^2}+12 I_T\left(\sqrt{m_F^2+\delta m^2}\right)+12 \sumint_K G(K)\frac{\delta \Pi(K)}{K^2+m_F^2+\delta m^2}\,,\nonumber\\
  \delta \Pi(K)&=&8 \sumint_Q G(K-Q)\left[\frac{2 \Sigma(Q)}{1+2 \Sigma(Q)}\Gamma(K,Q-K)-\delta \Gamma(K,Q-K)\right]\,.
\end{eqnarray}
The pressure in the R3 approximation is thus given by (\ref{eq:zform}) with $P_{\rm div}=\frac{m^4}{48}$ and
\begin{eqnarray}
  \label{eq:pr3}
  P_{R3}(T,\bar\mu^2)&=&-\frac{1}{2}\sumint_K \ln \frac{K^2+m_F^2+\delta m^2-\delta \Pi(K)}{K^2+m_F^2+\delta m^2}-\frac{1}{4}\sumint_K \delta \Pi(K) G(K)\nonumber\\
  &&-\frac{1}{2}\sumint_K \ln \left[1+2 \Sigma(K)\right]+\frac{1}{2}\sumint_K \Sigma(K) D(K)-\frac{1}{4}\sumint_K D(K)\delta \Sigma(K)\nonumber\\
  &&-J_T\left(\sqrt{m_F^2+\delta m^2}\right)+\frac{(m_F^2+\delta m^2)^2}{48}\nonumber\\
  &&-\frac{(m_F^2+\delta m^2)}{24}\left(m_F^2+\frac{3}{\pi}\ln \frac{m_F^2 e^1}{m_F^2+\delta m^2}\right)\,.
%
\end{eqnarray}
Note that again, the dependence on the renormalization scale $\bar \mu^2$ has dropped out in $P_{R3}$.

As with R2, results for $P_{R3}(T)$ can be obtained numerically using the same methods as those described in the appendix of Ref.~\cite{Romatschke:2019rjk}. $P_{R3}(T)$ is compared to $P_{R0}(T), P_{R1}(T)$, $P_{R2}(T)$ in Fig.~\ref{fig1}, confirming the notion that thermal masses, not the physics of thermal widths, constitutes the dominant physics for the overall magnitude of the thermal pressure. In the zero temperature limit, one finds $P_{R3}(T\rightarrow 0)= -0.021(1)$, which matches the known high-precision zero-temperature result at $g=\frac{m}{\lambda}=1$ from Refs.~\cite{Serone:2018gjo,Elias-Miro:2017xxf} upon converting to the present renormalization scheme:
\begin{equation}
  \label{eq:t0exact}
  P_{\rm ``exact''}(T=0)=0.0392(3)-\frac{m_F^4}{48}-\frac{m_F^2}{8\pi}=-0.0214(3).
  \end{equation}

\begin{figure}[t]
  \includegraphics[width=\linewidth]{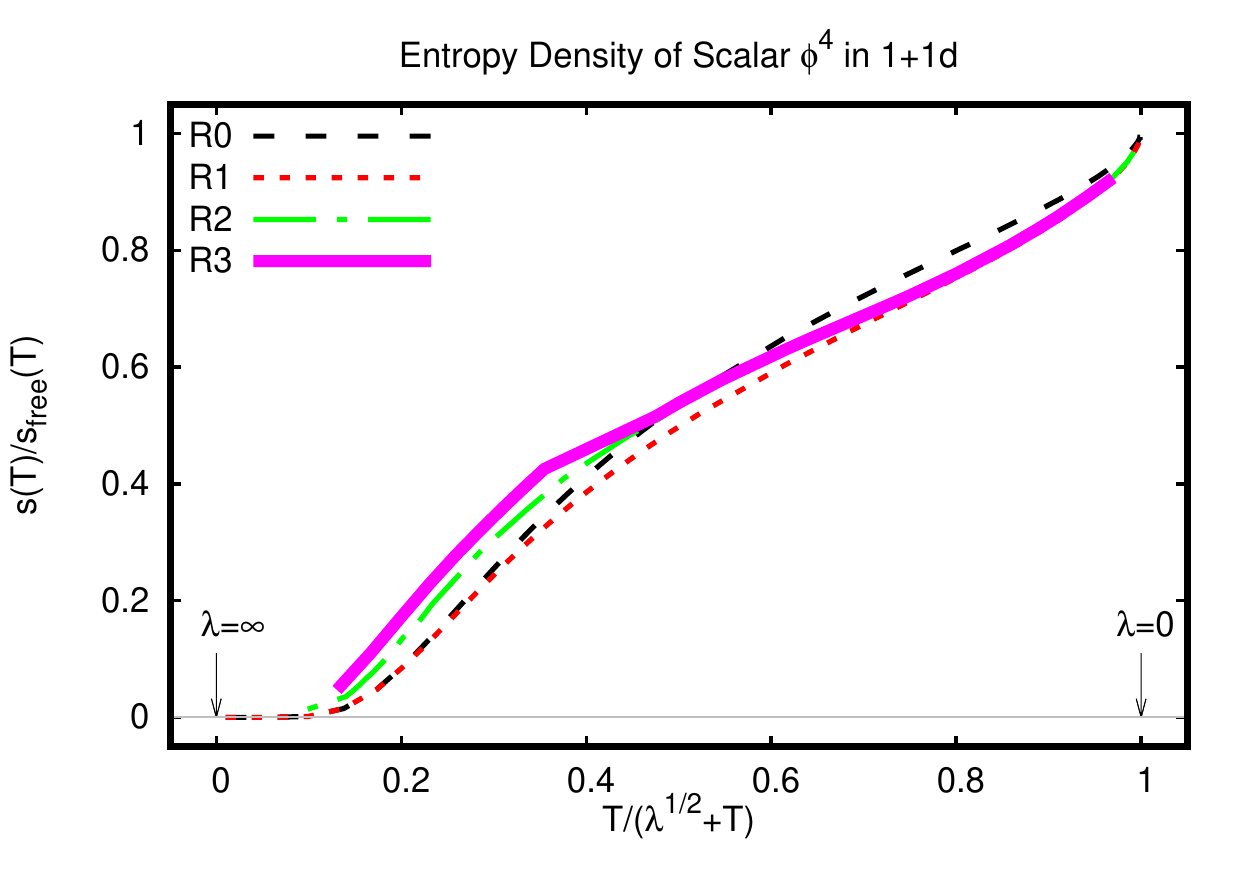}
  \caption{\label{fig2} Entropy density as a function of re-scaled temperature $T/\sqrt{\lambda}$ for $m_F^2=\lambda$ in the R0-R3 approximation scheme. Results do not depend on the renormalization scale choice $\bar\mu$. Horizontal plot axis is expressed as $\frac{T}{T+\sqrt{\lambda}}$ to allow compactification of the whole interval $T\in[0,\infty)$. Arrows indicate location of weak and strong coupling regime $\lambda=0$, $\lambda=\infty$, respectively.  Line thickness for R3 is representative for numerical errors arising from calculating numerical derivatives. The apparent ``kink'' in R3 at around $T\simeq 0.6$ is an artifact resulting from matching separately calculated values for the low and high temperature entropy, respectively. Full line is a guide to the eye.  See text for details.} 
\end{figure}

\section{Entropy density at finite temperature}

The entropy density $s\equiv \frac{dP}{dT}$ is readily calculated from the expressions for the thermal pressure given in Eqns.~(\ref{eq:pr0}), (\ref{eq:pr1}), respectively. One finds
\begin{equation}
s_{R0}(T), s_{R1}(T)=-\left.\frac{\partial J_T(M)}{\partial T}\right|_{M}=-\frac{M^2}{\pi T} \sum_{n=1}^\infty K_2\left(\frac{n M}{T}\right)\,,
\end{equation}
where $M^2$ in the R0, R1 approximations is given in Eqns.~(\ref{eq:r0gap}), (\ref{eq:r1gap}), respectively. While it is possible that also the R2- and R3-level approximation for the entropy density admits a simple expression, in practice one can calculate $s$ by performing a numerical temperature derivative from the existing results for the thermal pressure\footnote{For R3, where summation over a large number of Matsubara frequencies is most costly, I have calculated the numerical derivative explicitly using the first 40 (60) non-vanishing Matsubara frequencies at high (low) temperature. The results shown for R3 have then be filtered by fitting a low order polynomial to the numerical entropy values in order to remove numerical noise for the low temperature region $T<0.6$ and the high temperature region $T>0.6$, respectively.} (\ref{eq:pr2}). Note that at low temperatures, taking the derivative becomes numerically more challenging, which is why results at very low temperature are not reported for R2, R3 here.

The results for the entropy density in the R0-R3 level approximations, normalized by the free entropy density result $s_{\rm free}=\frac{\pi T}{3}$ are shown in Fig.~\ref{fig2}. As one can see from this figure, there is not only qualitative, but even overall quantitative agreement for the entropy density for all temperatures/couplings in the R0-R3 level approximations. Since R2 and R3 include the physics of thermal widths (albeit with different numerical factors), the fact that R2, R3 are close to the R0, R1 results for the entropy density may be taken as strong indication that thermal widths constitute a subleading effect to thermodynamic observables at any coupling strength.

If this result was valid in higher dimensions than 1+1d, this would be remarkable; it would suggest that when the physics of thermal widths is completely ignored, the resulting approximation schemes (R0, R1) give values for thermodynamic quantities which are good to better than 20 percent for all interaction strengths. 

Leaving studies of this possibility for future work, given the approximate results for $s/s_{\rm free}$ for 1+1 dimensions shown in Fig.~\ref{fig2}, I predict  that exact calculations would give $s/s_{\rm free}>0.9$ for $T>21$ and $s/s_{\rm free}<0.2$ for $T<0.2$ for $m_F^2=1$. Based on the agreement between R0-R3, I expect these predictions to be robust.

Moreover, given that the R3 approximation was found to be quantitatively similar to high-precision results for scalar $\phi^4$ at zero temperature in Ref.~\cite{Romatschke:2019rjk}, I predict that R3 results to be a good quantitative approximation to the exact result for $m_F^2=1$ for arbitrary temperatures/coupling values.

\begin{figure}[t]
  \includegraphics[width=\linewidth]{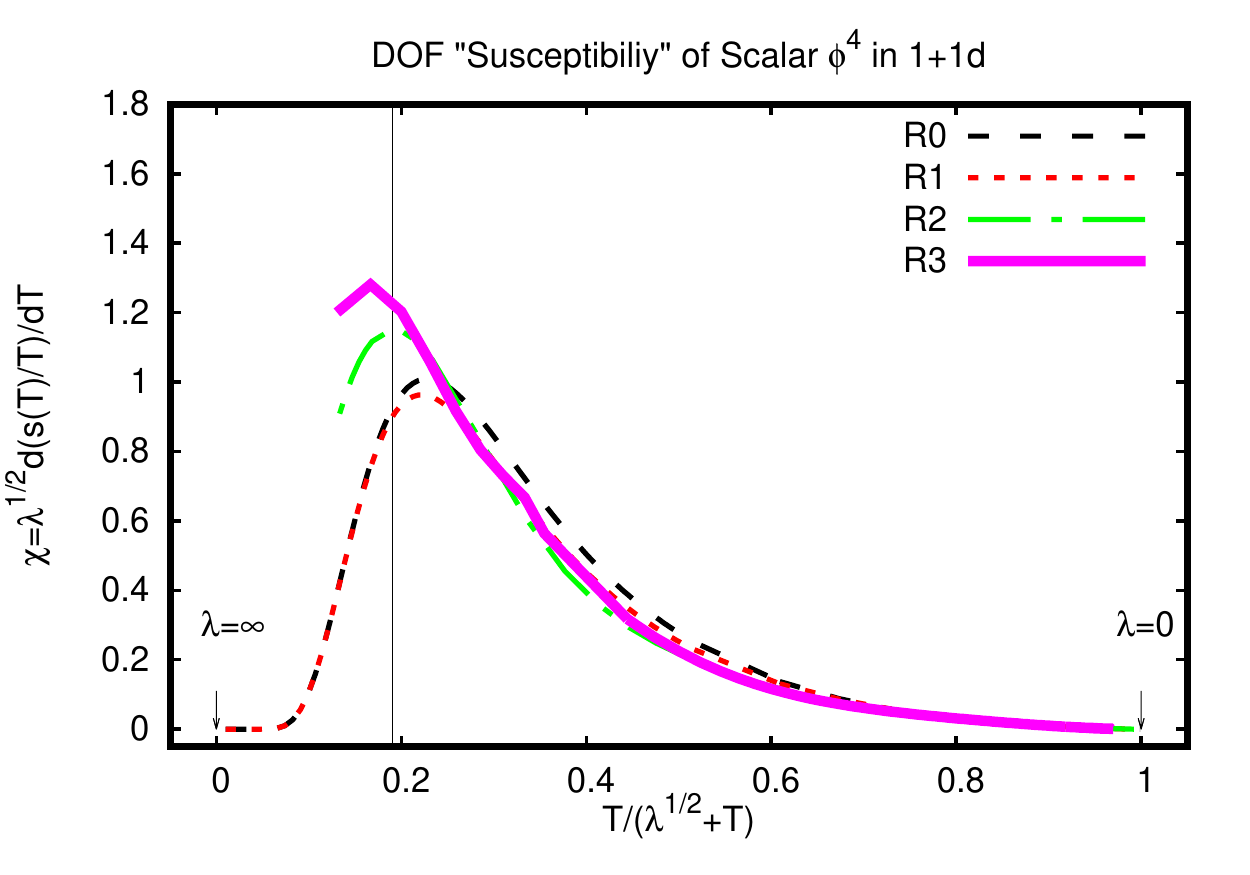}
  \caption{\label{fig4} ``Susceptibility'' for the effective number of degrees of freedom (\ref{eq:sus}) as a function of re-scaled temperature $T/\sqrt{\lambda}$ for $m_F^2=\lambda$ in the R0-R3 approximation scheme. Results do not depend on the renormalization scale choice $\bar\mu$. Horizontal plot axis is expressed as $\frac{T}{T+\sqrt{\lambda}}$ to allow compactification of the whole interval $T\in[0,\infty)$. Arrows indicate location of weak and strong coupling regime $\lambda=0$, $\lambda=\infty$, respectively. Line thickness for R3 is representative for numerical errors arising from calculating numerical derivatives. The apparent ``kink'' in R3 at around $T\simeq 0.6$ is an artifact resulting from matching separately calculated values for the low and high temperature entropy, respectively. Vertical line at $T\simeq 0.235$ is a guide to the eye. See text for details.} 
\end{figure}

\section{Cross-over transition between low and high temperature}

The behavior of the entropy density as a function of temperature, normalized to the free-field result as shown in Fig.~\ref{fig2},  bears similarities with that of full QCD in that there is a low-temperature regime where $s\simeq 0$ and a high-temperature regime where the entropy density approaches $s_{\rm free}$ from below. In QCD, the change in the normalized entropy density is associated with the change in the number of degrees of freedom from the confined low-temperature hadronic phase to the deconfined high-temperature quark-gluon plasma phase. Unlike pure Yang-Mills, the transition between confined and deconfined phase in full QCD with physical quark masses is known to be an analytic cross-over transition from lattice Monte-Carlo simulations \cite{Aoki:2006we}.

In the absence of a true transition, there is no true order parameter characterizing the low- and high-temperature phase. However, approximate order parameters such as the effective number of degrees of freedom given by
\begin{equation}
  {\rm DOF}(T)=\frac{s(T)}{s_{\rm free}}\,,
\end{equation}
in practice allow one to distinguish between the two phases.
The location of the cross-over transition between low- and high- temperature ``phase'' may therefore be estimated from the peak of the ``susceptibility''
\begin{equation}
  \label{eq:sus}
  \chi(T)\equiv \frac{d}{dT} {\rm DOF}(T)\,.
  \end{equation}

While there is no confinement mechanism operating in scalar field theory, one may nevertheless evaluate $\chi(T)$ for the R0-R3 approximations in order to distinguish between a low-temperature ``phase'' where $s\simeq 0$ and a high-temperature ``phase'' where $s\rightarrow s_{\rm free}$. The corresponding plot is shown in Fig.~\ref{fig4}, indicating a broad cross-over transition at $T\simeq 0.235$ for $m_F^2=1$.

\section{Speed of Sound at finite temperature}

\begin{figure}[t]
  \includegraphics[width=\linewidth]{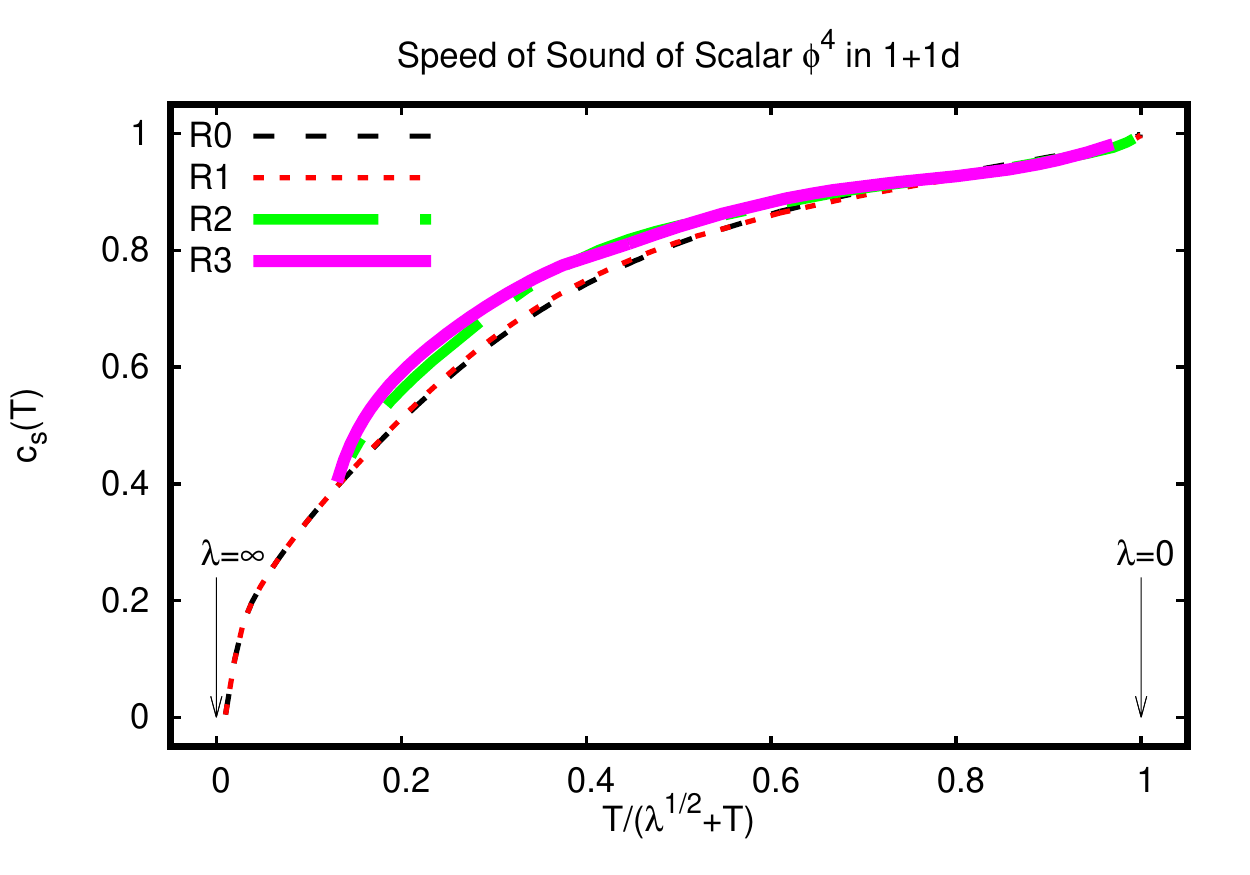}
  \caption{\label{fig3} Speed of sound $c_s(T)$ as a function of re-scaled temperature $T/\sqrt{\lambda}$ for $m_F^2=\lambda$ in the R0-R3 approximation scheme. Results do not depend on the renormalization scale choice $\bar\mu$. Horizontal plot axis is expressed as $\frac{T}{T+\sqrt{\lambda}}$ to allow compactification of the whole interval $T\in[0,\infty)$. Arrows indicate location of weak and strong coupling regime $\lambda=0$, $\lambda=\infty$, respectively. Line thickness for R2- and R3-level results is representative for numerical errors arising from calculating numerical derivatives. The apparent ``kink'' in R3 at around $T\simeq 0.6$ is an artifact resulting from matching separately calculated values for the low and high temperature entropy, respectively.   See text for details.} 
\end{figure}

The speed of sound $c_s(T)$ at finite temperature is calculated from the thermodynamic relation
\begin{equation}
  c_s(T)\equiv \sqrt{\frac{d \epsilon}{d P}}=\sqrt{\frac{s/T}{\frac{ds}{dT}}}\,,
\end{equation}
where $\epsilon=sT-P$ is the energy density. The corresponding derivative of the entropy density is readily calculated numerically and one finds the speed of sound in the R0-R3 level approximation shown in Fig.~\ref{fig3}. Similar to what has been found for the entropy density, the relative differences between R0-R3 for $c_s(T)$ are small for all temperatures/couplings shown. (Note that taking the second derivative of the pressure is numerically more difficult at low temperatures, which is why results for $c_s(T)$ are not reported for the lowest temperatures for R2, R3.)

\section{Summary and Conclusions}

In this work, I calculated thermodynamic properties for scalar $\phi^4$ in 1+1 dimensions at all temperatures/coupling values based on the non-perturbative resummation schemes R0-R3 developed in Ref.~\cite{Romatschke:2019rjk}.

Results found for the thermal pressure, entropy density and speed of sound in the R0-R3 scheme were found to be well-behaved for arbitrary temperature and coupling strength. Furthermore, the dependence on the renormalization scale choice \hbox{$\bar\mu\in [\pi T,4 \pi T]$} dropped out explicitly for observables calculated in R0-R3. Moreover, results obtained in the R0-R3 schemes were also found to be numerically close at all temperatures/ coupling values.
When contrasted with usual perturbation theory, these findings strongly suggest that the R0-R3 schemes are capable of providing quantitatively useful results for scalar $\phi^4$ theory even in the full non-perturbative regime.

In addition, the rapid rise of the entropy density as a function of temperature found in the R0-R3 calculations for $m_F^2=1$ hints at the possibility of an analytic cross-over between a low temperature and high temperature ``phase'' in scalar $\phi^4$ theory.

Given that scalar $\phi^4$ theory is amenable to direct numerical simulation using Monte Carlo lattice field theory techniques, I would encourage calculation of these thermodynamic observables on the lattice in the future.

  \section{Acknowledgments}

  This work was supported by the Department of Energy, DOE award No DE-SC0017905. I would like to thank P.~Bedaque and M.~Pinto for helpful discussions.

\bibliography{lambda}
\end{document}